\documentclass{article}
\usepackage{spconf,amsmath,graphicx,hyperref,bm}
\hypersetup{hidelinks}

\title{CNN-Transformer with Sigmoid Label for High-Resolution Sub-Degree DOA Estimation}

\name{\href{https://orcid.org/0009-0005-4597-0611}{Lifeng Wang}$^{\star}$, \href{https://orcid.org/0009-0005-1411-0520}{Xintong Li}$^{\dagger}$, and \href{https://orcid.org/0000-0002-9317-8013}{Aifei Liu}$^{\star}$\thanks{This work was supported by the XJTLU Research Development Funding (Grant No. RDF-25-01-053).}}
\address{$^{\star}$Academy of AI and Advanced Technology, Xi'an Jiaotong-Liverpool University, Taicang 215400, China\\
$^{\dagger}$Department of Electrical and Computer Engineering, Western University, London, ON N6A 3K7, Canada}

\begin{document}
\ninept
\maketitle

\begin{abstract}
This paper addresses the grid quantization error of deep-learning (DL)-based direction-of-arrival (DOA) estimation. Classification-based DOA methods discretize the continuous angular space into a grid, so that with a peak-based read-out, the accuracy and resolution are limited by the grid step and by the width of the label peak. To address this issue, we propose a \emph{sigmoid label} that encodes the sources' DOAs as the sum of angle-shifted monotonic sigmoid functions. Differentiating the output of a CNN-Transformer network trained with the sigmoid label produces sharp spectral peaks at the source DOAs. Combined with a moderately refined $0.25^\circ$ grid step, the CNN-Transformer with the proposed sigmoid label attains sub-degree accuracy and a $1^\circ$ resolution, breaking the limit of the grid quantization error. Under the unified CNN-Transformer backbone, we compare the proposed Sigmoid label with the Gaussian label, the 0-1 hard label, and the binary cross-entropy (BCE) baseline. Numerical results demonstrate that the Sigmoid label attains the highest accuracy and superior resolution compared to other labels. 

\end{abstract}

\begin{keywords}
Array signal processing, Direction of arrival (DOA) estimation, Deep learning, Sigmoid label, Quantization error.
\end{keywords}

\section{Introduction}
\label{sec:intro}
Direction of Arrival (DOA) estimation is fundamental to radar, wireless communication, and sonar \cite{i-1}, \cite{i-2}, \cite{i-3}, \cite{i-4}. Classical methods range from CBF \cite{i-6}, whose resolution is limited by the array aperture, to the high-resolution subspace methods MUSIC \cite{schmidt1986multiple} and ESPRIT \cite{roy1986estimation}, whose performance degrades at low SNRs and for closely spaced sources. Deep learning (DL) has recently brought remarkable robustness to DOA estimation, ranging from fully-connected deep neural networks (DNN) and convolutional neural networks (CNN) \cite{dl-1}, \cite{dl-2}, \cite{dl-3}, \cite{dl-4} to the local-global feature (LG)-CNN \cite{dl-5} and the hybrid CNN-Vision Transformer (ViT) model with multiple class tokens (named HMC-ViT) \cite{dl-6}.

In DL-based DOA estimation, the angular space is discretized into a grid, and the network outputs a spatial spectrum that is decoded by locating its peaks. This paradigm suffers from grid quantization error: because the peak is localized to a grid point, the accuracy is bounded by the grid size even for a perfectly trained network. Soft labels such as the Gaussian label have been adopted to smooth the target \cite{he2018deep,chen2024sdoanet}, yet the peak-based readout still confines the accuracy to the grid.

We propose a \emph{Sigmoid label} that encodes the sources' DOAs as the sum of monotonic sigmoid functions. Differentiating the predicted Sigmoid function yields exceptionally sharp spectral peaks at the sigmoid centers, whose positions directly give the source angles. Because these derivative peaks are far sharper and better separated than the Gaussian bump, they can be located reliably even for closely spaced sources. Combined with a moderate $0.25^\circ$ grid size, the proposed readout attains sub-degree resolution. Under the unified HMC-ViT backbone, we compare the Sigmoid label with a mean squared error (MSE) loss against the Gaussian label with an MSE loss, the 0-1 hard label (shortened as hard label) with an MSE loss, and the hard label with a binary cross-entropy (BCE) loss.

The main contributions are summarized as follows.
\begin{itemize}
\item A Sigmoid label that encodes multiple source angles as the sum of multiple monotonic sigmoid functions, which is easy to learn by DL-based methods and is fundamentally different from the existing peak-shaped labels.
\item A derivative readout specially designed for the Sigmoid label, which recovers the source angles at sub-degree resolution, breaking the $1^\circ$ quantization limit.
\item A comparison under a unified backbone showing that the Sigmoid label with an MSE loss outperforms the Gaussian label with an MSE loss, the hard label with an MSE loss, and the hard label with a BCE loss.
\end{itemize}

\section{Background}
\subsection{Array received signal}
A uniform linear array (ULA) of $M$ sensors receives $K$ far-field narrowband signals. Their directions of arrival (DOAs) are stacked in $\boldsymbol{\theta} = [\theta_1, \ldots, \theta_K]^T$, where $\theta_k$ refers to the DOA of the $k$-th signal. The snapshot observed at time $t$ reads
\begin{equation}
\mathbf{x}(t) = \sum \limits_{k=1}^{K} \mathbf{a}(\theta_k)s_k(t)+\mathbf{n}(t)\label{eq1},
\end{equation}
in which $\mathbf{x}(t) = [x_1(t), \ldots, x_M(t)]^T$ collects the $M$ element outputs, $\mathbf{n}(t)= [n_1(t), \ldots, n_M(t)]^T$ is zero-mean Gaussian noise with per-element variance $\sigma_n^2$, and $\mathbf{a}(\theta_k)$ is the steering vector
\begin{multline}
\mathbf{a}(\theta_k) = [1,e^{-j2\pi \frac{d_2}{\lambda} \sin(\theta_k)}, \ldots, e^{-j2\pi \frac{d_{M}}{\lambda} \sin(\theta_k)} ]^T,\label{eq2}
\end{multline}
Here $(\cdot)^T$ is the transpose, $j=\sqrt{-1}$ the imaginary unit, $d_i$ the position of the $i$-th sensor relative to the reference element, and $\lambda$ the carrier wavelength. Arranging the steering vectors as the columns of a matrix, \eqref{eq1} admits the compact form
\begin{equation}
\mathbf{x}(t) = \mathbf{A}(\theta)\mathbf{s}(t) + \mathbf{n}(t),\label{eq3}
\end{equation}
where $\mathbf{A}(\theta)=[\mathbf{a}(\theta_1), \mathbf{a}(\theta_2), \ldots, \mathbf{a}(\theta_K)]$, of size $M \times K$, collects the steering vectors and $\mathbf{s}(t)=[s_1(t),s_2(t),\ldots,s_K(t)]^T$ holds the source samples. Averaging over $N$ snapshots yields the sample covariance matrix
\begin{equation}\label{eq:R_est}
\widehat{\mathbf{R}} = \frac{1}{N} \sum\limits_{t=1}^{N} \mathbf{x}(t)\mathbf{x}^H(t),
\end{equation}
where $(\cdot)^H$ denotes the Hermitian transpose. The matrix $\widehat{\mathbf{R}}$ is used as the input to the CNN-Transformer introduced below.

\subsection{HMC-ViT}
As the unified backbone, the proposed method employs the hybrid multi-class token vision transformer (HMC-ViT) of \cite{dl-6}. HMC-ViT merges a convolutional neural network (CNN), which extracts local features, with a vision transformer (ViT) whose multiple class tokens each focus on a sub-region of the angular space, giving both low computational cost and strong spatial representation. We keep this backbone architecture unchanged and concentrate on the label design together with the readout strategy, which form the core of this work.

The proposed Sigmoid label and the Gaussian label are discretized with $0.25^\circ$ steps ($480$ grid points), while the 0-1 hard and BCE labels use the conventional $1^\circ$ grid ($121$ grid points).

\section{Problem statement}\label{sec:method}
In DL-based DOA estimation, the continuous angular space is first discretized into a grid with a fixed step $\rho_{in}$, and the network is trained to output a spatial spectrum $\mathbf{y} \in [0,1]^{G}$ over the $G$ grid points. The estimated angles are then recovered by locating the peaks of the output spectrum through a peak-searching procedure. This classification-like paradigm is shared by most existing DL-based DOA methods.

Two label designs dominate the existing methods, both of which form a \emph{peak} around each source. The 0-1 hard label assigns $1$ to the grid point associated with each true angle and $0$ elsewhere. The Gaussian label \cite{he2018deep,chen2024sdoanet} smooths this assignment by spreading a Gaussian-shaped bump around the true angle. Despite their different shapes, both labels are peak-shaped, and both are decoded by the same peak-based readout that returns the index of the grid point with the largest response. The 0-1 hard label can be trained with either the mean squared error (MSE) or the binary cross-entropy (BCE) loss, but both variants share the same peak-based readout and are therefore equally confined to the grid step.

The fundamental limitation of this paradigm lies in the peak-based readout. Because the peak-searching procedure returns the index of a discrete grid point, the estimated angle is always confined to the grid $\{\theta_g\}$, even when the network is perfectly trained. Consequently, the accuracy of the grid-confined peak-based readout is bounded by the grid step: for a $1^\circ$ grid, the quantization error of a single source is distributed in $[-\rho_{in}/2, \rho_{in}/2]$, where $\rho_{in} = 1^\circ$, thus limiting the accuracy of any label that is decoded by locating its peaks on the grid.

It is worth emphasizing that this quantization limit is \emph{structural} rather than a consequence of insufficient training. Both the 0-1 hard and the Gaussian labels are peak-shaped, and the peak-based readout can only localize a peak to the nearest grid point. A straightforward remedy is to refine the grid, e.g., reducing the grid size from $1^\circ$ to $0.1^\circ$; however, this increases the number of classes by a factor of $10$, which dramatically raises the difficulty of the classification task, the required network capacity, and the amount of training data. Most importantly, for the hard label, each output grid corresponds to one angle for generating training data. This implies that the training dataset correspondingly increases. For example, with a field of view of $[-60^\circ, 60^\circ]$ and a grid size of $1^\circ$, the number of angle combinations is $C_{121}^2=7260$ \cite{Aifei2026}. For a grid size of $0.1^\circ$, the number of angle combinations is $C_{1210}^2=731445$, increasing by a factor of $100$. This large data volume leads to significant training costs.

Motivated by the above observation, we argue that the bottleneck is not merely the grid step but the label design and its associated readout. In the next section, we propose a Sigmoid label whose derivative produces exceptionally sharp spectral peaks at the source angles. This attains sub-degree resolution without the prohibitive class-count growth of a naive $0.1^\circ$ refinement for the training dataset.

\section{Proposed Sigmoid label and its derivative}
\subsection{Sigmoid label design}
We distinguish the input-side grid step $\rho_{in}$, i.e., the spacing of the candidate source angles used to synthesize the training data, from the output-side grid step $\rho_{out}$, i.e., the spacing of the label vector. For the Sigmoid and Gaussian labels, $\rho_{in}=1^\circ$ and $\rho_{out}=0.25^\circ$, whereas for the 0-1 hard and BCE labels the two coincide, $\rho_{in}=\rho_{out}=1^\circ$. For the soft-label design, we discretize the angular range $[\theta_{\min}, \theta_{\max}]$ into a grid with step $\rho_{out}$, yielding the grid angles $\theta_g = \theta_{\min} + g\,\rho_{out}$ for $g = 0, 1, \ldots, G-1$, where $G = \frac{\theta_{\max}-\theta_{\min}}{\rho_{out}} = 480$.

For a scenario with $K$ sources located at $\{\theta_1, \ldots, \theta_K\}$, the Sigmoid label is denoted as a vector $\mathbf{y}$, and its $g$-th element $y_g$ is constructed as a normalized sum of angle-shifted sigmoid functions:
\begin{equation}
    y_g = \frac{1}{K} \sum_{k=1}^{K} \sigma\!\left( \frac{\theta_g - \theta_k}{w} \right), \quad g = 0, 1, \ldots, G-1,
\end{equation}
where $\sigma(\cdot)$ denotes the logistic sigmoid function $\sigma(x) = 1/(1 + e^{-x})$, and $w = 0.2^\circ$ is a smoothing width, in degrees, that controls the sharpness of each sigmoid.

The Sigmoid label is a monotonically increasing function of the angle. Each source contributes one sigmoid function that rises from $0$ to $1$ around its true angle, and the $K$ sigmoids accumulate into a Sigmoid-like cumulative distribution function (CDF) over $[0, 1]$. This is fundamentally different from the Gaussian label and the 0-1 hard label, both of which form a peak around each source.

\begin{figure}[!htbp]
    \centering
    \includegraphics[width=\columnwidth]{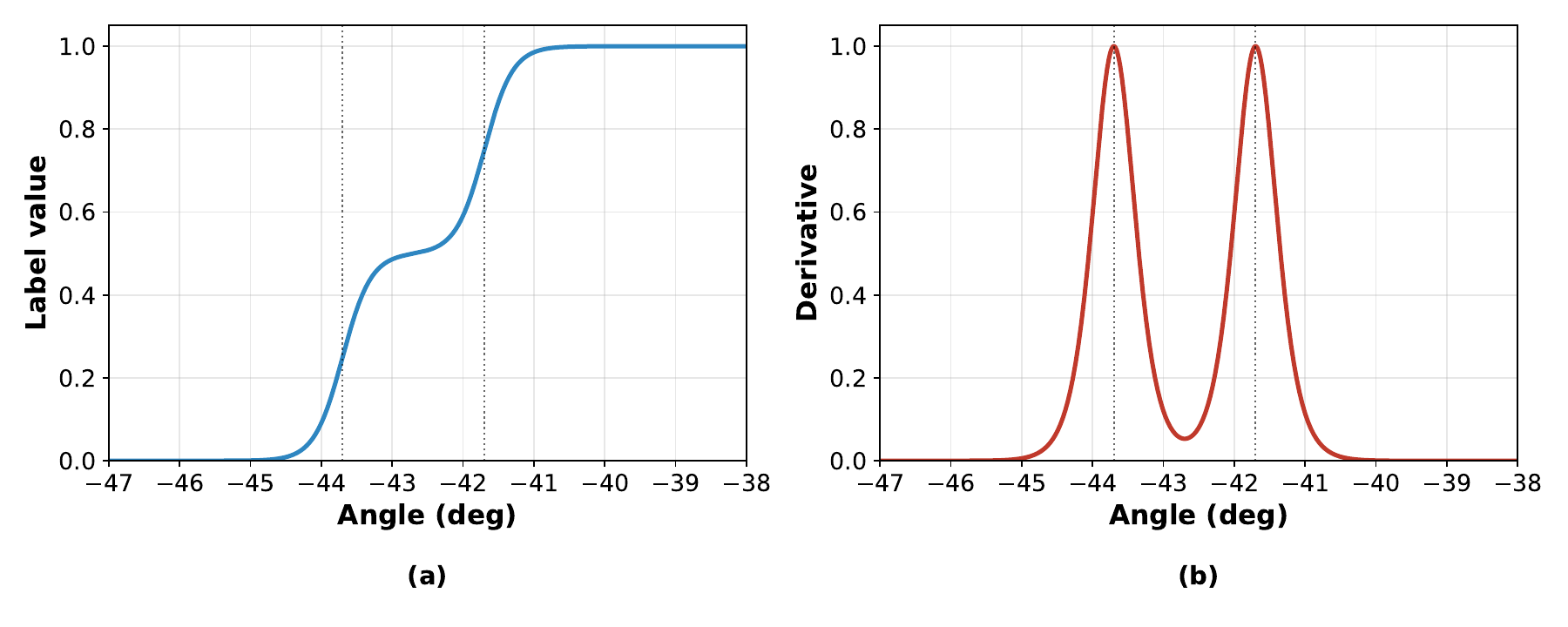}
    \caption{The proposed Sigmoid label and its first derivative for two sources at $-43.7^\circ$ and $-41.7^\circ$ (separation $2^\circ$, $w=0.2^\circ$). (a) The Sigmoid label (before derivative) forms a monotonic CDF. (b) The first derivative (after derivative) forms two sharp peaks at the source angles.}
    \label{fig:deriv}
\end{figure}

Fig.~\ref{fig:deriv} illustrates the proposed label and its derivative for two sources located at $-43.7^\circ$ and $-41.7^\circ$, i.e., separated by $2^\circ$, with $w=0.2^\circ$. In Fig.~\ref{fig:deriv}(a), the Sigmoid label forms a monotonically increasing curve, where each source contributes one sigmoid function and the two sigmoids accumulate into an S-shaped cumulative distribution function. Taking the first derivative of this curve, as shown in Fig.~\ref{fig:deriv}(b), converts each sigmoid into a single sharp peak located exactly at its center, i.e., at the true source angle. Even at a $2^\circ$ separation, the two peaks remain clearly separated, providing the basis for the derivative readout below.

Since $\sigma(0) = 0.5$, each sigmoid reaches its mid-point at the corresponding true angle $\theta_k$. Therefore, the source angles can be recovered from the positions of the sigmoid centers, rather than from isolated peaks. Differentiating a sigmoid yields a single sharp peak located exactly at its center, and this motivates the derivative readout described next, which recovers the source angles.

Since the Sigmoid label is bounded within $[0,1]$, a sigmoid activation is applied at the network output so that the prediction matches the range of the target cumulative distribution function. Beyond bounding the output, this activation also acts as a smooth nonlinearity that suppresses large oscillations of the network output, keeping the predicted curve smooth and monotonically increasing, so that its first derivative remains a well-defined, unimodal peak at each sigmoid.

\subsection{First-derivative-based readout}
The Sigmoid label is read out by differentiating the network output and locating its peaks. Let $\hat{\mathbf{y}}$ denote the network output; its first derivative with respect to the angle is computed by central differences,
\begin{equation}
    d_g = \frac{\hat{y}_{g+1} - \hat{y}_{g-1}}{2\rho_{out}}, \quad g = 1, 2, \ldots, G-2,
\end{equation}
where $\hat{y}_g$ denotes the $g$-th element of $\hat{\mathbf{y}}$, i.e., the network output at the grid angle $\theta_g = \theta_{\min} + g\,\rho_{out}$ (so $\hat{y}_{g+1}$ and $\hat{y}_{g-1}$ are the outputs at the two neighbouring grid angles $\theta_{g+1}$ and $\theta_{g-1}$). The source angles are recovered from the local maxima of $\mathbf{d} = \{d_g\}$ through
\begin{equation}
    \hat{\theta} = \theta_{\min} + g_p\,\rho_{out},
\end{equation}
where $g_p$ is the grid index of a detected peak; the $K$ peaks with the largest heights are retained and mapped to angles through the grid. A standard monotonicity-preserving smoothing is applied to the network output before the differentiation in order to suppress non-monotonic fluctuations.

The key advantage of the Sigmoid representation lies in the shape of these derivative peaks. For a Gaussian-shaped label, the spectrum itself is a broad bump whose peak must be located by an argmax, and the resolution is limited by the grid step. In contrast, the Sigmoid derivative forms peaks whose width is governed by the sigmoid transition width $w$ (in degrees) rather than by the grid step, so that each peak is markedly narrower and taller than the Gaussian bump. Consequently, two closely spaced sources yield two clearly separated derivative peaks located at the true source angles, whereas the two sub-peaks of the Gaussian spectrum are broad and biased away from the true angles, so that its located angles carry a large estimation error; this mechanism enables sub-degree resolution on the $0.25^\circ$ grid.

\section{Simulation Results}\label{sec:results}
In this section, we evaluate the proposed Sigmoid label under a unified CNN-Transformer backbone, and compare it with the Gaussian label, the 0-1 hard label, and the BCE baseline.

\subsection{Simulation Setup}
We consider a uniform linear array (ULA) of $M=16$ elements with half-wavelength spacing, i.e., the element spacing is $d=\lambda/2$ with the wavelength $\lambda=0.3$~m. $K$ far-field narrowband sources impinge on the array from directions within the range $[-60^\circ, 60^\circ]$. The element spacing, the wavelength, and the array configuration are fixed throughout all experiments.

Two grid settings are adopted. The proposed Sigmoid label and the Gaussian label are discretized with a step of $\rho_{out}=0.25^\circ$, giving $G=480$ grid points over $[-60^\circ, 60^\circ)$, while the 0-1 hard and BCE labels follow the conventional $1^\circ$ grid with $G=121$ grid points. For reference, the classical MUSIC spectrum is evaluated on the same $0.25^\circ$ grid ($480$ points). The sigmoid transition width is set to $w=0.2^\circ$, which corresponds to a $5\%$--$95\%$ transition band of $1.18^\circ$ and a derivative-peak full width at half maximum (FWHM) of $0.70^\circ$. For the Gaussian label, the standard deviation of the Gaussian bump is set to $\sigma=2^\circ$.

The sources are assumed to have equal power, and the additive noise is spatially and temporally white Gaussian noise; the signal-to-noise ratio (SNR) is defined as the ratio of the source power to the noise power at each array element. The network input is the upper-triangular part (excluding the diagonal) of the estimated covariance matrix $\widehat{\mathbf{R}}$, organized into three channels corresponding to its real part, imaginary part, and phase, which gives a $3\times120$ input to the CNN-Transformer. The HMC-ViT backbone employs $8$ class tokens that jointly attend to the spatial features. The synthesized dataset is randomly split into $85\%$ for training and $15\%$ for validation.

For training, the data are synthesized for two sources ($K=2$) at five SNRs, i.e., $\{-15,-10,-5,0,5\}$~dB, with the number of snapshots $N=1000$. The number of angle combinations is $C_{121}^{2}=7260$; with $10$ realizations per combination and per SNR, the training set contains $7260 \times 5 \times 10 = 363{,}000$ samples. The network is trained for $500$ epochs using the Adam optimizer with an initial learning rate of $1\times10^{-4}$ and a batch size of $64$. The mean squared error (MSE) loss is used for the proposed Sigmoid label, the Gaussian label, and the 0-1 hard label, while the binary cross-entropy (BCE) loss is used for the BCE baseline, following the original HMC-ViT configuration.

For testing, all the methods are evaluated on the \emph{same} dataset under the same conditions, so that the comparison is fair. The RMSE is evaluated at SNRs from $-20$ to $5$~dB for the accuracy study, and the probability of resolution is evaluated at $0$ and $5$~dB; both metrics are computed over $M_c=1000$ Monte Carlo trials with $N=1000$ snapshots each, unless otherwise stated. The RMSE is defined as
\begin{equation}
    \mathrm{RMSE} = \sqrt{\frac{1}{M_c K}\sum_{i=1}^{M_c}\sum_{k=1}^{K} (\hat{\theta}_{k,i} - \theta_k)^2},
\end{equation}
where $\hat{\theta}_{k,i}$ is the estimate of $\theta_k$ in the $i$-th Monte Carlo trial.

\subsection{Spatial Spectra}
\begin{figure}[!htbp]
    \centering
    \includegraphics[width=\columnwidth]{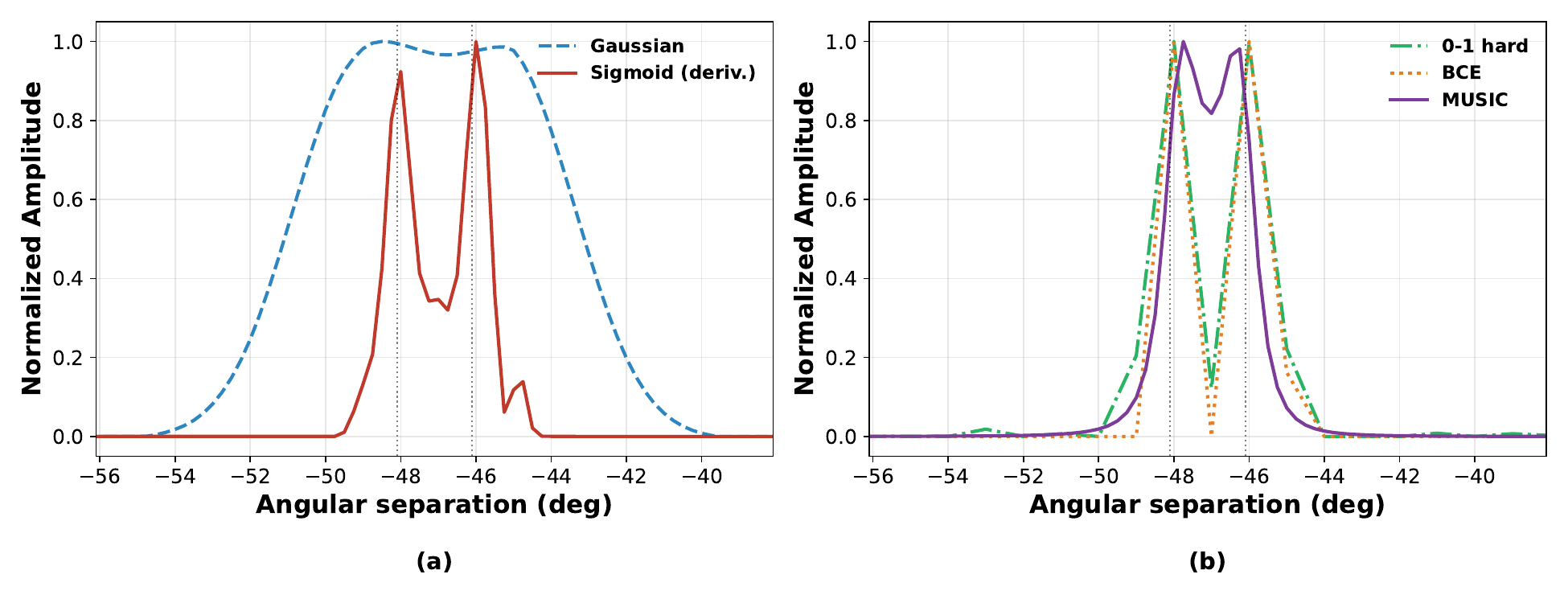}
    \caption{Output spectra at a $2^\circ$ separation (SNR $=5$~dB) for two sources at $-48.1^\circ$ and $-46.1^\circ$: (a) the Gaussian and Sigmoid-derivative spectra; (b) the 0-1 hard, BCE and MUSIC spectra.}
    \label{fig:spectra}
\end{figure}

We first inspect the output spectra at a small ($2^\circ$) separation, at an SNR of $5$~dB, to illustrate the mechanism. The two sources are located at $-48.1^\circ$ and $-46.1^\circ$. As shown in Fig.~\ref{fig:spectra}(a), the Gaussian spectrum forms a broad bump with two poorly separated sub-peaks that are biased away from the true source angles, so that the sources cannot be located accurately; the Sigmoid derivative, by contrast, produces two sharp and well-separated peaks located exactly at the sources. Fig.~\ref{fig:spectra}(b) further shows the 0-1 hard, BCE and MUSIC spectra, which also contain two peaks but are much broader and carry higher side lobes than the Sigmoid derivative; the narrow, side-lobe-free derivative peaks are what make the Sigmoid estimates fall on the true source angles. This demonstrates that differentiating the Sigmoid output converts the source angles into sharp, unbiased spectral peaks, which is the mechanism enabling reliable sub-degree resolution.

\subsection{Performance versus SNR}
\begin{figure}[!htbp]
    \centering
    \includegraphics[width=\columnwidth]{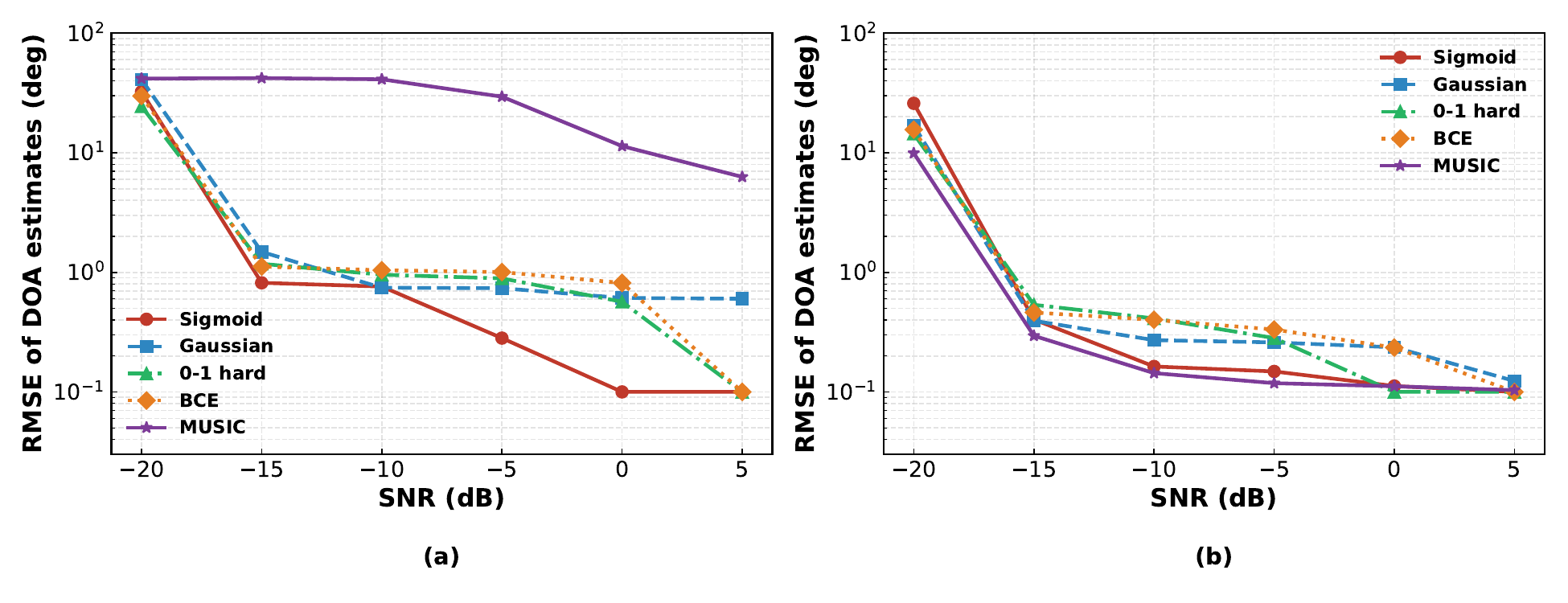}
    \caption{RMSE versus SNR of the proposed Sigmoid label, the three baselines and MUSIC, for a small ($2^\circ$, left) and a large ($20^\circ$, right) angular separation.}
    \label{fig:rmse_snr_labels}
\end{figure}

We compare the proposed Sigmoid label with the three baselines and the classical MUSIC algorithm. Fig.~\ref{fig:rmse_snr_labels} shows the RMSE versus SNR for a small ($2^\circ$) and a large ($20^\circ$) angular separation. For the large separation (Fig.~\ref{fig:rmse_snr_labels}(b)), all methods perform well at high SNR, and the classical MUSIC algorithm attains an RMSE close to the proposed method, since well-separated sources are favorable for subspace-based methods. The situation changes drastically at the small separation (Fig.~\ref{fig:rmse_snr_labels}(a)): MUSIC fails completely, with an RMSE of tens of degrees, whereas the proposed Sigmoid label attains the lowest RMSE among all the methods (about $0.2^\circ$ at high SNR), far below the $1^\circ$ grid quantization floor. The Gaussian, 0-1 hard and BCE labels remain at a much higher level, because their detected peaks are broad and, especially for the Gaussian label, biased away from the true angles; this bias grows as the separation shrinks. This shows that the advantage of the proposed method lies in the small-separation regime, where the subspace method and the peak-shaped labels fail, while at large separations all the methods are adequate and the difference is small.

\subsection{Angular Resolution}
We evaluate the ability to resolve two closely spaced sources. The first source is fixed at $-48.1^\circ$ and the second is placed at a separation from $1^\circ$ to $10^\circ$, over $1000$ Monte Carlo trials. Consistent with the readout described above, a trial is counted as \emph{resolved} when the two strongest peaks of the spectrum lie on opposite sides of the true midpoint and the sum of their absolute angle errors is smaller than the true separation.

\begin{figure}[!htbp]
    \centering
    \includegraphics[width=\columnwidth]{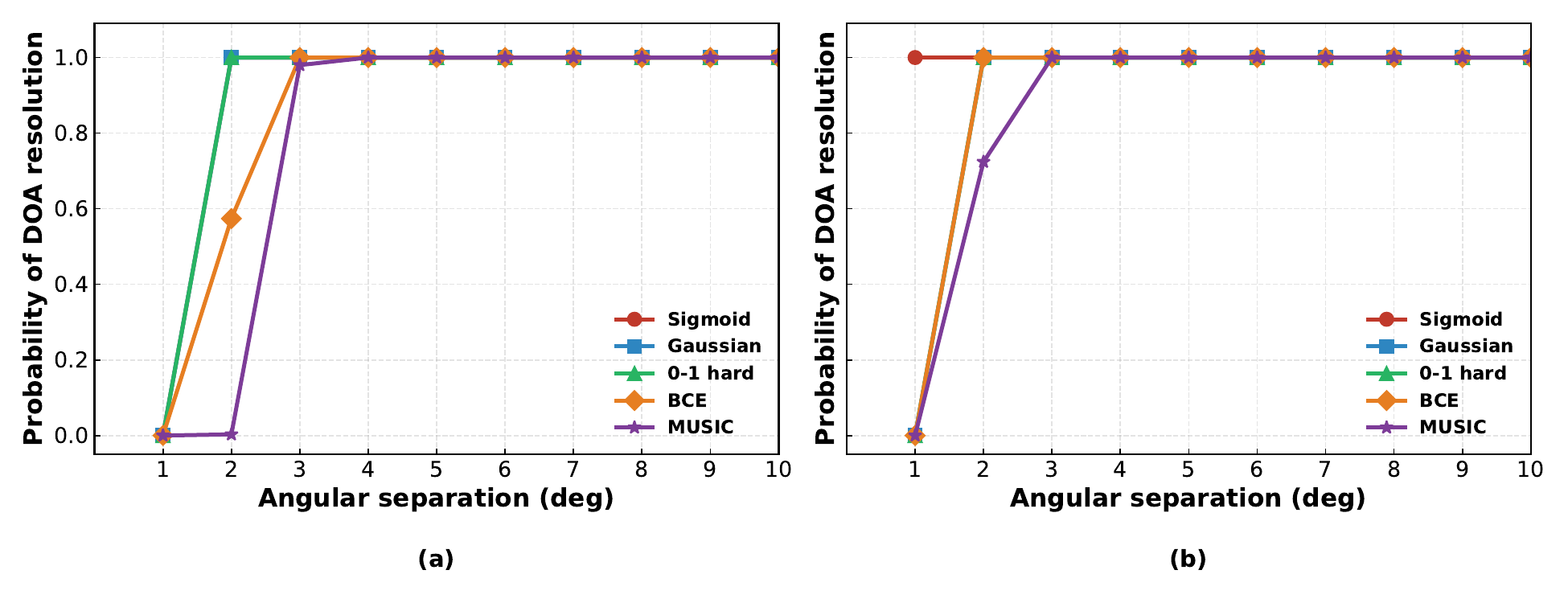}
    \caption{Probability of resolution versus angular separation of the proposed Sigmoid label, the three baselines, and MUSIC, at an SNR of $0$~dB (left) and $5$~dB (right).}
    \label{fig:resolution}
\end{figure}

Fig.~\ref{fig:resolution} shows the probability of resolution versus the angular separation for the two SNRs of $0$ and $5$~dB. At $5$~dB (Fig.~\ref{fig:resolution}(b)), the proposed Sigmoid label resolves two sources separated by only $1^\circ$ with probability $1$, whereas the Gaussian and 0-1 hard labels, the BCE baseline, and MUSIC all fail to resolve the $1^\circ$ separation; MUSIC additionally requires at least $2^\circ$--$3^\circ$ before it begins to resolve the sources. At the lower SNR of $0$~dB (Fig.~\ref{fig:resolution}(a)), none of the methods resolve the $1^\circ$ separation; from $2^\circ$ onward the Sigmoid, Gaussian and 0-1 hard labels all begin to resolve the sources, whereas the BCE and MUSIC spectra require a larger separation. Overall, the Sigmoid label is the only method that resolves the $1^\circ$ separation (at $5$~dB), exhibiting the best resolution capability at small separations.

\section{Conclusion}
In this paper, we proposed a Sigmoid label for deep-learning-based DOA estimation. Different from the conventional Gaussian-shaped and 0-1 hard labels that form a peak around each source, the Sigmoid label encodes the source angles as a monotonic cumulative distribution function composed of sigmoid functions. Combined with a readout that takes the derivative of the predicted label, turning the smooth sigmoid into sharp peaks located at the true source DOAs, the proposed method recovers the source angles at sub-degree resolution, breaking the $1^\circ$ quantization limit of the peak-based decoding. Under a unified CNN-Transformer backbone, the Sigmoid label was compared with the Gaussian label, the 0-1 hard label and the BCE baseline in terms of the output spectra, the estimation accuracy versus the SNR, and the angular resolution. Simulation results demonstrated that the Sigmoid label attains higher accuracy and better resolution. In particular, at an SNR of $5$~dB, the CNN-Transformer with the sigmoid label uniquely resolves two sources separated by only $1^\circ$, where the ones with other labels and the classical MUSIC fail.

\vfill\pagebreak

\bibliographystyle{IEEEbib}
\bibliography{cite}

\end{document}